\documentclass[11pt]{article}
\usepackage{amssymb,amsmath,amsfonts}
\usepackage{graphicx}
\usepackage{graphics}
\usepackage{eepic,epsfig}

\begin{document}

\title{Polarization of the fermionic vacuum by a global monopole with finite
core}
\author{E. R. Bezerra de Mello$^{1}$\thanks{%
E-mail: emello@fisica.ufpb.br}\, and A. A. Saharian$^{1,2}$\thanks{%
E-mail: saharian@ictp.it} \\
\\
\textit{$^1$Departamento de F\'{\i}sica-CCEN, Universidade Federal da Para%
\'{\i}ba}\\
\textit{58.059-970, Caixa Postal 5.008, Jo\~{a}o Pessoa, PB, Brazil}\vspace{%
0.3cm}\\
\textit{$^2$Department of Physics, Yerevan State University,}\\
\textit{375025 Yerevan, Armenia}}
\maketitle

\begin{abstract}
We study the vacuum polarization effects associated with a massive fermionic
field in a spacetime produced by a global monopole considering a nontrivial
inner structure for it. In the general case of the spherically symmetric
static core with finite support we evaluate the vacuum expectation values of
the energy-momentum tensor and the fermionic condensate in the region
outside the core. These quantities are presented as the sum of point-like
global monopole and core-induced contributions. The asymptotic behavior of
the core-induced vacuum densities are investigated at large distances from
the core, near the core and for small values of the solid angle
corresponding to strong gravitational fields. As an application of general
results the flower-pot model for the monopole's core is considered and the
expectation values inside the core are evaluated.
\end{abstract}

\bigskip

{PACS number(s): 03.07.+k, 98.80.Cq, 11.27.+d}

\newpage

\section{Introduction}

Symmetry braking phase transitions in the early universe have several
cosmological consequences and provide an important link between particle
physics and cosmology. In particular, different types of topological objects
may have been formed by the vacuum phase transitions after Planck time \cite%
{Kibble,V-S}. These include domain walls, cosmic strings and monopoles. A
global monopole is a spherical symmetric gravitational topological defect
created by a phase transition of a system comprised by self-coupling scalar
field, $\varphi ^{a}$, whose original global $O(3)$ symmetry is
spontaneously broken to $U(1)$. The matter fields play the role of an order
parameter which outside the monopole's core acquires a non-vanishing value.
The global monopole was first introduced by Sokolov and Starobinsky \cite%
{Soko77}. A few years later, the gravitational effects associated with a
global monopole have been considered in Ref. \cite{B-V}, where the authors
have found that for points far from the monopole's center, the geometry is
similar to the black-hole with a solid angle deficit. Neglecting the mass
term we get the point-like global monopole spacetime with the metric tensor
given by the following line element
\begin{equation}
ds^{2}=dt^{2}-dr^{2}-\alpha ^{2}r^{2}(d\theta ^{2}+\sin ^{2}\theta d\phi
^{2})\ ,  \label{mmetric}
\end{equation}%
where the parameter $\alpha ^{2}$ is smaller than unity and depends on the
energy scale where the symmetry is broken. It is of interest to note that
the effective metric produced in superfluid $^{3}\mathrm{He-A}$ by a
monopole is described by the line element (\ref{mmetric}) with the negative
angle deficit, $\alpha >1$, which corresponds to the negative mass of the
topological object \cite{Volo98}. The quasiparticles in this model are
chiral and massless fermions.

In quantum field theory the non-trivial topology of the global monopole
spacetime induces non-zero vacuum expectation values for physical
observables. The quantum effects due to the point-like global monopole
spacetime on the matter fields have been considered for massless scalar \cite%
{M-L} and fermionic \cite{EVN} fields, respectively. The influence of the
non-zero temperature on these polarization effects has been discussed in
Ref. \cite{C-E} for scalar and fermionic fields. Moreover, the calculation
of quantum effects on massless scalar field in a higher dimensional global
monopole spacetime has also been developed in Ref. \cite{E}. The quantum
effects of a scalar field induced by a composite topological defect
consisting a cosmic string on a $p$-dimensional brane and a $(m+1)$%
-dimensional global monopole in the transverse extra dimensions are
investigated in Ref. \cite{Beze06Comp}. The combined vacuum polarization
effects by the non-trivial geometry of a global monopole and boundary
conditions imposed on the matter fields are investigated as well. In this
direction, the total Casimir energy associated with massive scalar field
inside a spherical region in the global monopole background has been
analyzed in Refs. \cite{MKS,EVN1} by using the zeta function regularization
procedure. Scalar Casimir densities induced by spherical boundaries have
been calculated in Refs. \cite{A-M,Saha03mon} to higher dimensional global
monopole spacetime by making use of the generalized Abel-Plana summation
formula \cite{Sahrev}. More recently, using also this formalism, a similar
analysis for spinor fields obeying MIT bag boundary conditions has been
developed in Refs.~\cite{Saha-Mello,Beze06}.

In general, the quantum analysis of matter fields in global monopole
spacetime consider this object as been a point-like one. Because of this
fact the renormalized vacuum expectation value of the energy-momentum tensor
presents singularities at the monopole's center. Of course, this kind of
problem cannot be expected in a realistic model. So a procedure to cure this
divergence is to consider the global monopole as having a non-trivial inner
structure. In fact, in a realistic point of view, the global monopoles have
a characteristic core radius determined by the symmetry braking energy scale
where the symmetry of the system is spontaneously broken. A simplified model
for the monopole core is presented in Ref. \cite{Hara90}. In this model the
region inside the core is described by the de Sitter geometry. The vacuum
polarization effects associated with a massless scalar field in the region
outside the core of this model are investigated in Ref. \cite{Spin05}. In
particular, it has been shown that long-range effects can take place due to
the non-trivial core structure. Recently the quantum analysis of a scalar
field in a higher-dimensional global monopole spacetime considering a
general spherically symmetric model for the core, has been considered in
Ref. \cite{S-E}. In the four-dimensional version of this model the induced
electrostatic self-energy and self-force for a charged particle are
investigated in Ref. \cite{Beze06Ind}. Continuing in this direction, in the
present paper we analyze the effects of global monopole core on properties
of the fermionic quantum vacuum. The most important quantities
characterizing these properties are the vacuum expectation values of the
energy-momentum tensor and the fermionic condensate. Though the
corresponding operators are local, due to the global nature of the vacuum,
the vacuum expectation values describe the global properties of the bulk and
carry an important information about the structure of the defect core. In
addition to describing the physical structure of the quantum field at a
given point, the energy-momentum tensor acts as the source of gravity in the
Einstein equations. It therefore plays an important role in modelling a
self-consistent dynamics involving the gravitational field. The problem
under consideration is also of separate interest as an example with
gravitational polarization of the fermionic vacuum, where all calculations
can be performed in a closed form. The corresponding results specify the
conditions under which we can ignore the details of the interior structure
and approximate the effect of the global monopole by the idealized model.
The exactly solvable models of this type are useful for testing the validity
of various approximations used to deal with more complicated geometries, in
particular in black-hole spacetimes.

The plan of this paper is as follows. Section \ref{sec:EigFunc} presents the
geometry of the problem and the eigenfunctions for a massive spinor field.
By using these eigenfunctions, in Section \ref{sec:ExtReg} we evaluate the
vacuum expectation values of the energy-momentum tensor and the fermionic
condensate in the region outside the monopole core. In Section \ref%
{sec:flowerpot} we consider the special case of the flower-pot model for the
core and the vacuum expectation values are investigated in both exterior and
interior regions. Various limiting cases are considered. The main results of
the paper are summarized in Section \ref{sec:Conc}. In Appendix \ref%
{sec:app1} we discuss the contribution of possible bound states into the
vacuum expectation values of the energy-momentum tensor and show that in the
flower-pot model there are no bound states. Throughout we use the system of
units with $\hbar =c=1$.

\section{Model and the eigenfunctions for the spinor field}

\label{sec:EigFunc}

In this section we analyze the eigenfunctions for a massive fermionic field
in background of the global monopole geometry considering a nontrivial inner
structure to the latter. The explicit expression for the metric tensor in
the region inside the monopole core is unknown and here we consider a
general model for a four-dimensional global monopole with a core of radius $%
a $, assuming that the geometry of the spacetime is described by two
distinct metric tensors in the regions inside and outside the core. Adopting
this model we can learn under which conditions we can ignore the specific
details for its core and consider the monopole as being a point-like object.
In spherical coordinates we will consider the corresponding line element in
the interior region, $r<a$, with the form
\begin{equation}
ds^{2}=e^{2u(r)}dt^{2}-e^{2v(r)}dr^{2}-e^{2h(r)}(d\theta ^{2}+\sin
^{2}\theta d\phi ^{2})\ .  \label{metricinside}
\end{equation}%
In the region outside, $r>a$, the metric tensor is given by the line element
(\ref{mmetric}), where the parameter $\alpha $ codifies the presence of the
global monopole. For an idealized point-like global monopole the geometry is
described by line element (\ref{mmetric}) for all values of the radial
coordinate and it has singularity at the origin.

In Eq. (\ref{metricinside}) the functions $u(r)$, $v(r)$, $h(r)$ are
continuous at the core boundary, consequently they satisfy the conditions
\begin{equation}
u(a)=v(a)=0,\ h(a)=\ln (\alpha a)\ .  \label{uvbound}
\end{equation}%
If there is no surface energy-momentum tensor on the bounding surface $r=a$,
the radial derivatives of these functions are continuous as well. When the
surface energy-momentum tensor is present the junctions in the radial
derivatives of the components of the metric tensor are expressed in terms of
the surface energy density and stresses\footnote{%
This point will be analyzed in more detail at the end of the next section}.
Introducing a new radial coordinate $\tilde{r}=e^{h(r)}$ with the core
center at $\tilde{r}=0$, the angular part of line element (\ref{metricinside}%
) is written in the standard Minkowskian form. However, with this choice, in
general, we will obtain non-standard form of the angular part in the
exterior line element (\ref{mmetric}). In the model under consideration we
will assume that inside the core the spacetime geometry is regular. In
particular, from the regularity of the interior geometry at the core center
one has the conditions $u(r),v(r)\rightarrow 0$ and $h(r)\sim \ln \tilde{r}$
for $\tilde{r}\rightarrow 0$.

We are interested in quantum effects of a spinor field propagating on
background of the spacetime described by line elements (\ref{mmetric}) and (%
\ref{metricinside}). The dynamics of the massive spinor field in curved
spacetime is governed by the Dirac equation
\begin{equation}
i\gamma ^{\mu }\nabla _{\mu }\psi -M\psi =0\ ,  \label{Direq}
\end{equation}%
with the covariant derivative operator
\begin{equation}
\nabla _{\mu }=\partial _{\mu }+\Gamma _{\mu }\ .  \label{Covder}
\end{equation}%
Here $\gamma ^{\mu }=e_{(a)}^{\mu }\gamma ^{(a)}$ are the Dirac matrices in
curved spacetime, and $\Gamma _{\mu }$ is the spin connection given in terms
of the flat $\gamma $ matrices by the relation
\begin{equation}
\Gamma _{\mu }=\frac{1}{4}\gamma ^{(a)}\gamma ^{(b)}e_{(a)}^{\nu }e_{(b)\nu
;\mu }\ .  \label{Gammamu}
\end{equation}%
In the equations above $e_{(a)}^{\mu }$ is the vierbein satisfying the
condition $e_{(a)}^{\mu }e_{(b)}^{\nu }\eta ^{ab}=g^{\mu \nu }$.

In the region inside the monopole core we choose the basis tetrad
\begin{equation}
e_{(a)}^{\mu }=\left(
\begin{array}{cccc}
e^{-u(r)} & 0 & 0 & 0 \\
0 & e^{-v(r)}\sin \theta \cos \varphi  & e^{-h(r)}\cos \theta \cos \varphi
& -e^{-h(r)}\sin \varphi /\sin \theta  \\
0 & e^{-v(r)}\sin \theta \sin \varphi  & e^{-h(r)}\cos \theta \sin \varphi
& e^{-h(r)}\cos \varphi /\sin \theta  \\
0 & e^{-v(r)}\cos \theta  & -e^{-h(r)}\sin \theta  & 0%
\end{array}%
\right) \ ,  \label{e}
\end{equation}%
where the rows of the matrix are specified by the index $a$ and the columns
by the index $\mu $. By using Eq. (\ref{e}), for the non-vanishing
components of the spin connection we find
\begin{eqnarray}
\Gamma _{0} &=&\frac{1}{2}u^{\prime }e^{u-v}\gamma ^{(0)}{\vec{\gamma}}\cdot
{\hat{r}}\ ,  \notag \\
\Gamma _{2} &=&\frac{i}{2}(1-h^{\prime }e^{h-v}){\vec{\Sigma}}\cdot {\hat{%
\varphi}}\ ,  \label{ConComp} \\
\Gamma _{3} &=&-\frac{i}{2}\sin \theta (1-h^{\prime }e^{h-v}){\vec{\Sigma}}%
\cdot {\hat{\theta}}\ ,  \notag
\end{eqnarray}%
where the prime denotes derivative with respect to the radial coordinate, $%
\hat{r}\ ,\ \hat{\theta}$ and $\hat{\varphi}$ are the standard unit vectors
along the three spatial directions in spherical coordinates, and
\begin{equation}
\vec{\Sigma}=\left(
\begin{array}{cc}
\vec{\sigma} & 0 \\
0 & \vec{\sigma}%
\end{array}%
\right) \ ,  \label{Sigma}
\end{equation}%
with $\vec{\sigma}=(\sigma _{1},\sigma _{2},\sigma _{3})$ being the Pauli
matrices. From the obtained results, for the combination entering in the
Dirac equation we have
\begin{equation}
\gamma ^{\mu }\Gamma _{\mu }=\frac{u^{\prime }e^{-v}}{2}{\vec{\gamma}}\cdot {%
\hat{r}}-e^{-h}(1-h^{\prime }e^{h-v}){\vec{\gamma}}\cdot {\hat{r}}\ .
\label{gamGam}
\end{equation}%
After some intermediate steps, the Dirac equation reads
\begin{equation}
ie^{-u}\gamma ^{(0)}\partial _{t}\psi +ie^{-u/2-v-h}\gamma ^{(r)}\partial
_{r}(e^{u/2+h}\psi )-ie^{-h}\gamma ^{(r)}({\vec{\Sigma}}\cdot {\vec{L}}%
+1)\psi -M\psi =0\ ,  \label{Direq1}
\end{equation}%
with $\vec{L}$ being the standard angular momentum operator.

To find the vacuum expectation value (VEV) of the energy-momentum tensor we
need the corresponding eigenfunctions. For the problem under consideration
there are two types of eigenfunctions with different parities which we will
distinguish by the index $\sigma =0,1$. These functions are specified by the
total angular momentum $j=1/2,\ 3/2,\ ...$ and its projection $m=-j,\ -j+1,\
...,\ j$. Assuming the time dependence in the form $e^{-i\omega t}$, the
four-component spinor fields specified by the set of quantum numbers $\beta
=(\sigma kjm)$ with $k^{2}=\omega ^{2}-M^{2}$, can be written in terms of
two-component ones as
\begin{equation}
\psi _{\beta }=e^{-i\omega t}\left(
\begin{array}{c}
f_{\beta }(r)\Omega _{jl_{\sigma }m}(\theta ,\varphi ) \\
n_{\sigma }g_{\beta }(r)\Omega _{jl_{\sigma }^{\prime }m}(\theta ,\varphi )%
\end{array}%
\right) \ ,  \label{psisig}
\end{equation}%
where
\begin{equation}
n_{\sigma }=(-1)^{\sigma },\;l_{\sigma }=j-\frac{n_{\sigma }}{2},\;l_{\sigma
}^{\prime }=j+\frac{n_{\sigma }}{2},  \label{nsigma}
\end{equation}%
and $\Omega _{jl_{\sigma }m}(\theta ,\varphi )$ are the standard spinor
spherical harmonics (see, for instance, Ref. \cite{Bere82}). The latter are
eigenfunctions of the operator ${\mathcal{K}}={\vec{\sigma}}\cdot {\vec{L}}+I
$ as shown below:
\begin{equation}
{\mathcal{K}}\Omega _{jl_{\sigma }m}=-\kappa _{\sigma }\Omega _{jl_{\sigma
}m}\ ,\;\kappa _{\sigma }=-n_{\sigma }(j+1/2).  \label{SphHarm}
\end{equation}%
Note that we have the relation
\begin{equation}
\Omega _{jl_{\sigma }^{\prime }m}=i^{l_{\sigma }-l_{\sigma }^{\prime }}(\hat{%
n}\cdot \vec{\sigma})\Omega _{jl_{\sigma }m}\ .  \label{relsphhar}
\end{equation}

Substituting the function $\psi _{\beta }$ into the Dirac equation above,
and using for the flat Dirac matrices the representation given in Ref. \cite%
{Bere82}, for the radial functions we obtain a set of two coupled linear
differential equations:
\begin{eqnarray}
f_{\beta }^{\prime }+(u^{\prime }/2+h^{\prime }+\kappa _{\sigma
}e^{v-h})f_{\beta } &=&(M+e^{-u}\omega )e^{v}g_{\beta }\ ,  \label{f1} \\
g_{\beta }^{\prime }+(u^{\prime }/2+h^{\prime }-\kappa _{\sigma
}e^{v-h})g_{\beta } &=&(M-e^{-u}\omega )e^{v}f_{\beta }\ .  \label{g1}
\end{eqnarray}%
In the region $r>a$ for the radial functions we have the solutions%
\begin{eqnarray}
f_{\beta }(r) &=&f_{\beta }^{\mathrm{(ex)}}(r)=\frac{1}{\sqrt{r}}\left[
c_{1}J_{\nu _{\sigma }}(kr)+c_{2}Y_{\nu _{\sigma }}(kr)\right] ,\;
\label{fext} \\
g_{\beta }(r) &=&g_{\beta }^{\mathrm{(ex)}}(r)=-\frac{1}{\sqrt{r}}\frac{%
n_{\sigma }k}{\omega +M}\left[ c_{1}J_{\nu _{\sigma }+n_{\sigma
}}(kr)+c_{2}Y_{\nu _{\sigma }+n_{\sigma }}(kr)\right] ,  \label{gext}
\end{eqnarray}%
where $J_{\nu _{\sigma }}(x)$ and $Y_{\nu _{\sigma }}(x)$ are the Bessel and
Neumann functions of the order
\begin{equation}
\nu _{\sigma }=\frac{j+1/2}{\alpha }-\frac{n_{\sigma }}{2}.  \label{nusig}
\end{equation}%
By taking into account relation (\ref{relsphhar}), the corresponding
eigenfunctions are written in the form%
\begin{equation}
\psi _{\beta }^{\mathrm{(ex)}}=e^{-i\omega t}\left(
\begin{array}{c}
f_{\beta }^{\mathrm{(ex)}}(r)\Omega _{jl_{\sigma }m}(\theta ,\varphi ) \\
-ig_{\beta }^{\mathrm{(ex)}}(r)(\hat{n}\cdot \vec{\sigma})\Omega
_{jl_{\sigma }m}(\theta ,\varphi )%
\end{array}%
\right) \ .  \label{psiex}
\end{equation}%
In Eqs. (\ref{fext}) and (\ref{gext}), the integration constants $c_{1}$ and
$c_{2}$ are determined from the matching condition with the interior
solution.

The regular solution to Eqs. (\ref{f1}), (\ref{g1}) in the interior region, $%
r<a$, we will denote by%
\begin{equation}
f_{\beta }(r)=R_{1,n_{\sigma }}(r,k),\;g_{\beta }(r)=-\frac{n_{\sigma }k}{%
\omega +M}R_{2,n_{\sigma }}(r,k).  \label{fgint}
\end{equation}%
Near the core center, $\tilde{r}\rightarrow 0$, these functions behave like $%
R_{1,n_{\sigma }}\varpropto \tilde{r}^{j+(1-n_{\sigma })/2}$ and $%
R_{2,n_{\sigma }}\varpropto \tilde{r}^{j+(1+n_{\sigma })/2}$, where the
radial coordinate $\tilde{r}$ is introduced in the paragraph after formula (%
\ref{uvbound}). Note that the functions $R_{1,n_{\sigma }}(r,k)$ and $%
n_{\sigma }kR_{2,-n_{\sigma }}(r,-k)/(M-\omega )$ are solutions of the same
equation and, hence,
\begin{equation}
R_{2,-n_{\sigma }}(r,-k)=\mathrm{const}\cdot R_{1,n_{\sigma }}(r,k).
\label{relR21}
\end{equation}%
The interior eigenfunctions have the form%
\begin{equation}
\psi _{\beta }^{\mathrm{(in)}}=e^{-i\omega t}\left(
\begin{array}{c}
R_{1,n_{\sigma }}(r,k)\Omega _{jl_{\sigma }m}(\theta ,\varphi ) \\
\frac{in_{\sigma }k}{\omega +M}R_{2,n_{\sigma }}(r,k)(\hat{n}\cdot \vec{%
\sigma})\Omega _{jl_{\sigma }m}(\theta ,\varphi )%
\end{array}%
\right) \ .  \label{psibetin}
\end{equation}%
From the continuity condition of the eigenfunctions on the surface $r=a$,
for the coefficients $c_{1}$ and $c_{2}$ in Eq. (\ref{fext}) one finds%
\begin{eqnarray}
c_{1} &=&-\frac{\pi }{2}n_{\sigma }ka^{3/2}\left[ R_{1,n_{\sigma
}}(a,k)Y_{\nu _{\sigma }+n_{\sigma }}(ka)-R_{2,n_{\sigma }}(a,k)Y_{\nu
_{\sigma }}(ka)\right] ,  \label{c1} \\
c_{2} &=&\frac{\pi }{2}n_{\sigma }ka^{3/2}\left[ R_{1,n_{\sigma
}}(a,k)J_{\nu _{\sigma }+n_{\sigma }}(ka)-R_{2,n_{\sigma }}(a,k)J_{\nu
_{\sigma }}(ka)\right] .  \label{c2}
\end{eqnarray}%
Note that from Eqs. (\ref{f1}), (\ref{g1}), for the values of the interior
radial functions on the boundary of the core we have the following relations%
\begin{eqnarray}
-n_{\sigma }kR_{2,n_{\sigma }}(a,k) &=&R_{1,n_{\sigma }}^{\prime
}(a,k)+\left( u_{a}^{\prime }/2+h_{a}^{\prime }-n_{\sigma }\lambda /a\right)
R_{1,n_{\sigma }}(a,k),  \label{R21rel1} \\
n_{\sigma }kR_{1,n_{\sigma }}(a,k) &=&R_{2,n_{\sigma }}^{\prime
}(a,k)+\left( u_{a}^{\prime }/2+h_{a}^{\prime }+n_{\sigma }\lambda /a\right)
R_{2,n_{\sigma }}(a,k),  \label{R21rel2}
\end{eqnarray}%
where $R_{j,n_{\sigma }}^{\prime }(a,k)=\partial R_{j,n_{\sigma
}}(r,k)/\partial r|_{r=a}$, $u_{a}^{\prime }=du/dr|_{r=a}$, $h_{a}^{\prime
}=dh/dr|_{r=a}$, and we have introduced the notation%
\begin{equation}
\lambda =(j+1/2)/\alpha .  \label{nu}
\end{equation}

Substituting the expressions for the coefficients $c_{1}$ and $c_{2}$ into
the formulae for the radial eigenfunctions in the exterior region, one finds%
\begin{equation}
\psi _{\beta }^{\mathrm{(ex)}}=\frac{\pi R_{1,n_{\sigma }}(a,k)e^{-i\omega t}%
}{2\sqrt{r/a}}\left(
\begin{array}{c}
g_{\nu _{\sigma },0}(ka,kr)\Omega _{jlm} \\
\frac{in_{\sigma }k}{\omega +M}g_{\nu _{\sigma },n_{\sigma }}(ka,kr)(\hat{n}%
\cdot \vec{\sigma})\Omega _{jlm}%
\end{array}%
\right) ,  \label{psiint}
\end{equation}%
with the notation%
\begin{equation}
g_{\nu _{\sigma },p}(x,y)=\tilde{Y}_{\nu _{\sigma }}(x)J_{\nu _{\sigma
}+p}(y)-\tilde{J}_{\nu _{\sigma }}(x)Y_{\nu _{\sigma }+p}(y),\;p=0,n_{\sigma
}.  \label{gnumu2}
\end{equation}%
Here and in what follows, for a function $F_{\nu _{\sigma }}(x)$ with $F=J,Y$
we use the tilded notation defined by the formula%
\begin{eqnarray}
\tilde{F}_{\nu _{\sigma }}(ka) &=&-n_{\sigma }ka\left[ F_{\nu _{\sigma
}+n_{\sigma }}(ka)-F_{\nu _{\sigma }}(ka)R_{2,n_{\sigma
}}(a,k)/R_{1,n_{\sigma }}(a,k)\right]  \notag \\
&=&kaF_{\nu _{\sigma }}^{\prime }(ka)-\left[ a\frac{R_{1,n_{\sigma
}}^{\prime }(a,k)}{R_{1,n_{\sigma }}(a,k)}+\frac{1}{2}au_{a}^{\prime
}+ah_{a}^{\prime }-\frac{1}{2}\right] F_{\nu _{\sigma }}(ka).
\label{Fnutilde}
\end{eqnarray}%
In deriving the second form we have used the relation (\ref{R21rel1}) and
the recurrence formula for the cylindrical functions.

The eigenfunctions are normalized by the condition%
\begin{equation}
\int d^{3}x\sqrt{\gamma }\psi _{\beta }^{+}\psi _{\beta ^{\prime }}=\delta
_{\beta \beta ^{\prime }},  \label{normcond}
\end{equation}%
where $\gamma $ is the determinant for the spatial metric and $\delta
_{\beta \beta ^{\prime }}$ is understood as the Kronecker delta symbol for
the discrete components of the collective index $\beta $ and as the Dirac
delta function for the continuous ones. As the normalization integral is
divergent for $\beta =\beta ^{\prime }$, the main contribution comes from
large values $r$. We may replace the cylindrical functions with the argument
$kr$ in Eq. (\ref{psiint}) by the corresponding asymptotic expressions. In
this way one finds%
\begin{equation}
R_{1,n_{\sigma }}^{2}(a,k)=\frac{2k(M+\omega )}{\pi ^{2}\alpha ^{2}a\omega }%
\left[ \tilde{J}_{\nu _{\sigma }}^{2}(ka)+\tilde{Y}_{\nu _{\sigma }}^{2}(ka)%
\right] ^{-1}.  \label{R1a}
\end{equation}%
This relation determines the normalization coefficient for the interior
region. In addition to the modes with real $k$, modes with purely imaginary $%
k$ can be present. These modes correspond to the bound states. The
eigenfunctions for the bound states and their normalization are discussed in
Appendix \ref{sec:app1}.

\section{Vacuum expectation values in the exterior region}

\label{sec:ExtReg}

In this section we consider the VEVs for the energy-momentum tensor and the
fermionic condensate in the region outside the global monopole core. We
expand the field operator in terms of the complete set of eigenfunctions $%
\{\psi _{\beta }^{(+)},\psi _{\beta }^{(-)}\}$:
\begin{equation}
\hat{\psi}=\sum_{\beta }[\hat{a}_{\beta }\psi _{\beta }^{(+)}+\hat{b}_{\beta
}^{+}\psi _{\beta }^{(-)}],  \label{operatorexp}
\end{equation}%
where $\hat{a}_{\beta }$ is the annihilation operator for particles, and $%
\hat{b}_{\beta }^{+}$ is the creation operator for antiparticles, $\psi
_{\beta }^{(+)}=\psi _{\beta }$, for $\omega >0$ and $\psi _{\beta
}^{(-)}=\psi _{\beta }$, for $\omega <0$. In order to find the VEV for the
operator of the energy-momentum tensor, we substitute the expansion (\ref%
{operatorexp}) and the analog expansion for the operator $\hat{\bar{\psi}}$
into the corresponding expression for spinor fields,
\begin{equation}
T_{\mu \nu }\{\hat{\bar{\psi}},\hat{\psi}\}=\frac{i}{2}[\hat{\bar{\psi}}%
\gamma _{(\mu }\nabla _{\nu )}\hat{\psi}-(\nabla _{(\mu }\hat{\bar{\psi}}%
)\gamma _{\nu )}\hat{\psi}]\ .  \label{EMTform}
\end{equation}%
By making use of the standard anticommutation relations for the annihilation
and creation operators, for the VEV one finds the following mode-sum formula
\begin{equation}
\langle 0|T_{\mu \nu }|0\rangle =\sum_{\beta }T_{\mu \nu }\{\bar{\psi}%
_{\beta }^{(-)}(x),\psi _{\beta }^{(-)}(x)\}\ ,  \label{modesum}
\end{equation}%
where $|0\rangle $ is the amplitude for the corresponding vacuum state.

Substituting the eigenfunctions (\ref{psiint}) into the mode-sum formula (%
\ref{modesum}), using the formula
\begin{equation}
\sum_{m=-j}^{j}|\Omega _{jl_{\sigma }m}(\theta ,\varphi )|^{2}=\frac{2j+1}{%
4\pi },  \label{sumOm}
\end{equation}%
and relation (\ref{R1a}), the VEV of the energy-momentum tensor is presented
in the form (no summation over $\mu $)
\begin{equation}
\langle 0|T_{\mu }^{\nu }|0\rangle =\frac{\delta _{\mu }^{\nu }}{8\pi \alpha
^{2}a^{3}r}\sum_{j=1/2}^{\infty }(2j+1)\sum_{\sigma =0,1}\int_{0}^{\infty }dx%
\frac{f_{\sigma \nu _{\sigma }}^{(\mu )}\left[ x,g_{\nu _{\sigma },p}(x,xr/a)%
\right] }{\tilde{J}_{\nu _{\sigma }}^{2}(x)+\tilde{Y}_{\nu _{\sigma }}^{2}(x)%
}\ .  \label{qext}
\end{equation}%
In formula (\ref{qext}) we use the notations
\begin{eqnarray}
f_{\sigma \nu _{\sigma }}^{(0)}\left[ x,g_{\nu _{\sigma },p}(x,y)\right]
&=&-x\left[ (\sqrt{x^{2}+M^{2}a^{2}}-Ma)g_{\nu _{\sigma },0}^{2}(x,y)\right.
\   \notag \\
&&\left. +(\sqrt{x^{2}+M^{2}a^{2}}+Ma)g_{\nu _{\sigma },n_{\sigma }}^{2}(x,y)%
\right] ,  \label{fnueps} \\
f_{\sigma \nu _{\sigma }}^{(1)}\left[ x,g_{\nu _{\sigma },p}(x,y)\right] &=&%
\frac{x^{3}}{\sqrt{x^{2}+M^{2}a^{2}}}\left[ g_{\nu _{\sigma
},0}^{2}(x,y)+g_{\nu _{\sigma },n_{\sigma }}^{2}(x,y)\right] \ -2f_{\sigma
\nu _{\sigma }}^{(2)}\left[ x,g_{\nu _{\sigma },p}(x,y)\right] ,
\label{fnup} \\
f_{\sigma \nu _{\sigma }}^{(\mu )}\left[ x,g_{\nu _{\sigma },p}(x,y)\right]
&=&\frac{x^{3}(2\nu _{\sigma }+n_{\sigma })}{2y\sqrt{x^{2}+M^{2}a^{2}}}%
g_{\nu _{\sigma },0}(x,y)g_{\nu _{\sigma },n_{\sigma }}(x,y),\;\mu =2,3,
\label{fnupperp}
\end{eqnarray}%
and the function $g_{\nu _{\sigma },p}(x,y)$ is defined by Eq. (\ref{gnumu2}%
). The expression on the right of Eq. (\ref{qext}) is divergent. It may be
regularized introducing a cutoff function $\psi _{\mu }(\omega )$ with the
cutting parameter $\mu $ which makes the divergent expressions finite and
satisfies the condition $\psi _{\mu }(\omega )\rightarrow 1$ for $\mu
\rightarrow 0$. After the renormalization the cutoff function is removed by
taking the limit $\mu \rightarrow 0$. In the discussion below we will
implicitly assume that the corresponding expressions are regularized. The
parts in the VEVs induced by the non-trivial structure of the core are
finite and do not depend on the regularization scheme used.

To find the part in the VEV of the energy-momentum tensor induced by the
non-trivial core structure, we subtract the corresponding components for the
point-like monopole geometry. The latter are given by the expressions which
are obtained from Eq. (\ref{qext}) replacing the integrand by $f_{\sigma \nu
_{\sigma }}^{(\mu )}\left[ x,J_{\nu _{\sigma }}(xr/a)\right] $ (see Ref.
\cite{Saha-Mello}). In order to evaluate the corresponding difference we use
the relation
\begin{equation}
\frac{f_{\sigma \nu _{\sigma }}^{(\mu )}\left[ x,g_{\nu _{\sigma },p}(x,xr/a)%
\right] }{\tilde{J}_{\nu _{\sigma }}^{2}(x)+\tilde{Y}_{\nu _{\sigma }}^{2}(x)%
}-f_{\sigma \nu _{\sigma }}^{(\mu )}\left[ x,J_{\nu _{\sigma }}(xr/a)\right]
=-\frac{1}{2}\sum_{s=1,2}\frac{\tilde{J}_{\nu _{\sigma }}(x)}{\tilde{H}_{\nu
_{\sigma }}^{(s)}(x)}f_{\sigma \nu _{\sigma }}^{(\mu )}\left[ x,H_{\nu
_{\sigma }}^{(s)}(xr/a)\right] \ ,  \label{relf}
\end{equation}%
where $H_{\nu }^{(s)}(z)$, $s=1,2$, are the Hankel functions. This allows to
present the vacuum energy-momentum tensor components in the form%
\begin{equation}
\langle 0|T_{\mu }^{\nu }|0\rangle =\langle 0_{\mathrm{m}}|T_{\mu }^{\nu
}|0_{\mathrm{m}}\rangle +\langle T_{\mu }^{\nu }\rangle _{\mathrm{c}}\ ,
\label{Tdecomp}
\end{equation}%
where $\langle 0_{\mathrm{m}}|T_{\mu }^{\nu }|0_{\mathrm{m}}\rangle $ is the
VEV for the point-like monopole and the part (no summation over $\mu $)
\begin{equation}
\langle T_{\mu }^{\nu }\rangle _{\mathrm{c}}=\frac{-\delta _{\mu }^{\nu }}{%
16\pi \alpha ^{2}a^{3}r}\sum_{j=1/2}^{\infty }(2j+1)\sum_{\sigma
=0,1}\sum_{s=1,2}\int_{0}^{\infty }dx\frac{\tilde{J}_{\nu _{\sigma }}(x)}{%
\tilde{H}_{\nu _{\sigma }}^{(s)}(x)}f_{\sigma \nu _{\sigma }}^{(\mu )}\left[
x,H_{\nu _{\sigma }}^{(s)}(xr/a)\right] \ ,  \label{qrbout}
\end{equation}%
is induced by the non-trivial core structure. In formula (\ref{qrbout}), the
integrand of the $s=1$ ($s=2$) term exponentially decreases in the upper
(lower) half of the complex plane $x$. Consequently, we rotate the
integration contour on the right of this formula by the angle $\pi /2$ for $%
s=1$ and by the angle $-\pi /2$ for $s=2$. This leads to the representation%
\begin{eqnarray}
\langle T_{\mu }^{\nu }\rangle _{\mathrm{c}} &=&\frac{-i\delta _{\mu }^{\nu }%
}{16\pi \alpha ^{2}a^{3}r}\sum_{j=1/2}^{\infty }(2j+1)\sum_{\sigma
=0,1}\int_{0}^{\infty }dx\sum_{s=1,2}\eta _{s}  \notag \\
&&\times \frac{\tilde{J}_{\nu _{\sigma }}(e^{\eta _{s}\pi i/2}x)}{\tilde{H}%
_{\nu _{\sigma }}^{(s)}(e^{\eta _{s}\pi i/2}x)}f_{\sigma \nu _{\sigma
}}^{(\mu )}\left[ e^{\eta _{s}\pi i/2}x,H_{\nu _{\sigma }}^{(s)}(e^{\eta
_{s}\pi i/2}xr/a)\right] ,  \label{Tmunuc2}
\end{eqnarray}%
where $\eta _{s}=(-1)^{s+1}$.

First of all let us consider the part of the integral over the interval $%
(0,Ma)$. By using the standard properties of the Hankel functions it can be
easily seen that in this interval one has
\begin{equation}
e^{i\pi \lambda }f_{\sigma \nu _{\sigma }}^{(\mu )}\left[ e^{\pi
i/2}x,H_{\nu _{\sigma }}^{(1)}(e^{\pi i/2}xr/a)\right] =e^{-i\pi \lambda
}f_{\sigma \nu _{\sigma }}^{(\mu )}\left[ e^{-\pi i/2}x,H_{\nu _{\sigma
}}^{(2)}(e^{-\pi i/2}xr/a)\right] .  \label{fnusigrel1}
\end{equation}%
Further, from equations (\ref{f1}) and (\ref{g1}) it follows that the
interior solution can be written as $R_{1,n_{\sigma }}(r,k)=\mathrm{const}%
\cdot R(r,\omega )$. On the base of this observation for the combination
entering into the tilted notations in Eq. (\ref{Tmunuc2}) one finds%
\begin{equation}
\frac{R_{1,n_{\sigma }}^{\prime }(a,e^{\pi i/2}x/a)}{R_{1,n_{\sigma
}}(a,e^{\pi i/2}x/a)}=\frac{R_{1,n_{\sigma }}^{\prime }(a,e^{-\pi i/2}x/a)}{%
R_{1,n_{\sigma }}(a,e^{-\pi i/2}x/a)},\;0\leqslant x\leqslant Ma.
\label{RelR}
\end{equation}%
Now, it can be seen that for these values $x$ one has%
\begin{equation}
e^{-i\pi \lambda }\frac{\tilde{J}_{\nu _{\sigma }}(e^{\pi i/2}x)}{\tilde{H}%
_{\nu _{\sigma }}^{(1)}(e^{\pi i/2}x)}=e^{i\pi \lambda }\frac{\tilde{J}_{\nu
_{\sigma }}(e^{-\pi i/2}x)}{\tilde{H}_{\nu _{\sigma }}^{(2)}(e^{-\pi i/2}x)}.
\label{relJH}
\end{equation}%
Combining relations (\ref{fnusigrel1}) and (\ref{relJH}), we see that the
part of the integral in Eq. (\ref{Tmunuc2}) over the interval $(0,Ma)$
vanishes.

To simplify the part of the integral over the interval $(Ma,\infty )$, we
note that for the functions with different parities the following relation
takes place:%
\begin{equation}
e^{i\pi \lambda }f_{\sigma \nu _{\sigma }}^{(\mu )}\left[ e^{\pi
i/2}x,H_{\nu _{\sigma }}^{(1)}(e^{\pi i/2}xr/a)\right] =-e^{-i\pi \lambda
}f_{\sigma ^{\prime }\nu _{\sigma ^{\prime }}}^{(\mu )}\left[ e^{-\pi
i/2}x,H_{\nu _{\sigma ^{\prime }}}^{(2)}(e^{-\pi i/2}xr/a)\right] ,
\label{frel3}
\end{equation}%
where $\sigma ,\sigma ^{\prime }=0,1$, and $\sigma \neq \sigma ^{\prime }$.
Further we note that the functions $R_{2,n_{\sigma ^{\prime }}}(r,-k)$ and $%
R_{1,n_{\sigma }}(r,k)$ satisfy the same equation and, hence, $%
R_{2,n_{\sigma ^{\prime }}}(r,-k)=\mathrm{const}\cdot R_{1,n_{\sigma }}(r,k)$%
. By using relations (\ref{R21rel1}) and (\ref{R21rel2}), now it can be seen
that%
\begin{equation}
\frac{R_{2,n_{\sigma ^{\prime }}}(a,ke^{\mp i\pi /2})}{R_{1,n_{\sigma
^{\prime }}}(a,ke^{\mp i\pi /2})}=\frac{R_{1,n_{\sigma }}(a,ke^{\pm i\pi /2})%
}{R_{2,n_{\sigma }}(a,ke^{\pm i\pi /2})}.  \label{RelratR}
\end{equation}%
From this formula, by taking into account that $\nu _{\sigma ^{\prime }}=\nu
_{\sigma }+n_{\sigma }$, we find the relation%
\begin{equation}
e^{-i\pi \lambda }\frac{\tilde{J}_{\nu _{\sigma }}(e^{\pi i/2}x)}{\tilde{H}%
_{\nu _{\sigma }}^{(1)}(e^{\pi i/2}x)}=e^{i\pi \lambda }\frac{\tilde{J}_{\nu
_{\sigma ^{\prime }}}(e^{-\pi i/2}x)}{\tilde{H}_{\nu _{\sigma ^{\prime
}}}^{(2)}(e^{-\pi i/2}x)}.  \label{RelJH3}
\end{equation}%
Combining formulae (\ref{frel3}) and (\ref{RelJH3}), we see that the
different parities give the same contribution to the VEV of the
energy-momentum tensor.

By taking into account this result and introducing the modified Bessel
functions, for the core-induced part in the VEV we find (no summation over $%
\mu $)%
\begin{equation}
\langle T_{\mu }^{\nu }\rangle _{\mathrm{c}}=\frac{\delta _{\mu }^{\nu }}{%
2\pi ^{2}\alpha ^{2}r}\sum_{l=1}^{\infty }l\int_{M}^{\infty }dx\frac{x^{3}}{%
\sqrt{x^{2}-M^{2}}}\sum_{s=1,2}\frac{\bar{I}_{l/\alpha -1/2}^{(\eta
_{s},0)}(ax)}{\bar{K}_{l/\alpha -1/2}^{(\eta _{s},0)}(ax)}F_{l/\alpha
-1/2}^{(\mu ,\eta _{s})}\left[ x,K_{l/\alpha -1/2}(xr)\right] .
\label{Tmunu6}
\end{equation}%
Here for a given function $f_{\nu }(y)$ we use the notations%
\begin{eqnarray}
F_{\nu }^{(0,\eta _{s})}\left[ x,f_{\nu }(y)\right] &=&\left( \frac{M^{2}}{%
x^{2}}-1\right) \left[ \left( 1+\frac{i\eta _{s}M}{\sqrt{x^{2}-M^{2}}}%
\right) f_{\nu }^{2}(y)\right.  \notag \\
&&\left. -\left( 1-\frac{i\eta _{s}M}{\sqrt{x^{2}-M^{2}}}\right) f_{\nu
+1}^{2}(y)\right] ,  \label{Fnu0} \\
F_{\nu }^{(1,\eta _{s})}\left[ x,f_{\nu }(y)\right] &=&f_{\nu
}^{2}(y)-f_{\nu +1}^{2}(y)-\lambda _{f}\frac{2\nu +1}{y}f_{\nu }(y)f_{\nu
+1}(y),  \label{Fnu1} \\
F_{\nu }^{(\mu ,\eta _{s})}\left[ x,f_{\nu }(y)\right] &=&\lambda _{f}\frac{%
2\nu +1}{2y}f_{\nu }(y)f_{\nu +1}(y),\;\mu =2,3,  \label{Fnu2}
\end{eqnarray}%
with $\lambda _{K}=-1$ (the function $F_{\nu }^{(\mu ,\eta _{s})}\left[
x,I_{\nu }(y)\right] $ with $\lambda _{I}=1$ will be used below), and the
barred notation is defined by the formula%
\begin{equation}
\bar{f}^{(\eta _{s},\sigma )}(x)=xf^{\prime }(x)-\left[ a\frac{%
R_{1,n_{\sigma }}^{\prime }(a,e^{\eta _{s}\pi i/2}x/a)}{R_{1,n_{\sigma
}}(a,e^{\eta _{s}\pi i/2}x/a)}+au_{a}^{\prime }/2+ah_{a}^{\prime }-1/2\right]
f(x).  \label{fetasbar}
\end{equation}%
It can be checked that the core-induced part in the VEV of the
energy-momentum tensor obeys the continuity equation $\langle T_{\mu }^{\nu
}\rangle _{\mathrm{c};\nu }=0$, which for the geometry under consideration
takes the form%
\begin{equation}
r\frac{d}{dr}\langle T_{1}^{1}\rangle _{\mathrm{c}}+2\left( \langle
T_{1}^{1}\rangle _{\mathrm{c}}-\langle T_{2}^{2}\rangle _{\mathrm{c}}\right)
=0.  \label{conteq}
\end{equation}%
In addition, for a massless spinor field this part is traceless and the
trace anomaly is contained in the point-like monopole part only.

The fermionic condensate can be found from the formula for the VEV of the
energy-momentum tensor by making use of the relation $T_{\mu }^{\mu }=M\bar{%
\psi}\psi $. The condensate is presented in the form of the sum of idealized
point-like monopole and core-induced parts:%
\begin{equation}
\langle 0|\bar{\psi}\psi |0\rangle =\langle 0_{\mathrm{m}}|\bar{\psi}\psi
|0_{\mathrm{m}}\rangle +\langle \bar{\psi}\psi \rangle _{\mathrm{c}}.
\label{condens}
\end{equation}%
By taking into account formula (\ref{Tmunu6}) for the components of the
energy-momentum tensor, for the part coming from the non-trivial core
structure one finds%
\begin{eqnarray}
\langle \bar{\psi}\psi \rangle _{\mathrm{c}} &=&\frac{1}{2\pi ^{2}\alpha
^{2}r}\sum_{l=1}^{\infty }l\int_{M}^{\infty }dx\,x\sum_{s=1,2}\frac{\bar{I}%
_{l/\alpha -1/2}^{(\eta _{s},0)}(ax)}{\bar{K}_{l/\alpha -1/2}^{(\eta
_{s},0)}(ax)}  \notag \\
&&\times \left[ \left( \frac{M}{\sqrt{x^{2}-M^{2}}}-i\eta _{s}\right)
K_{l/\alpha -1/2}^{2}(xr)-\left( \frac{M}{\sqrt{x^{2}-M^{2}}}+i\eta
_{s}\right) K_{l/\alpha +1/2}^{2}(xr)\right] .  \label{condens1}
\end{eqnarray}%
This formula may also be derived directly from the mode-sum formula $\langle
0|\bar{\psi}\psi |0\rangle =\sum_{\beta }\bar{\psi}_{\beta }^{(-)}\psi
_{\beta }^{(-)}$ with the eigenfunctions (\ref{psiex}). In deriving formulae
(\ref{Tmunu6}) and (\ref{condens1}) we have assumed that no bound states
exist. In Appendix \ref{sec:app1} we show that these formulae are valid also
in the case when the bound states are present.

For $r>a$ the core-induced parts in the VEVs of the energy-momentum tensor
and the fermionic condensate, given by Eqs. (\ref{Tmunu6}) and (\ref%
{condens1}), are finite and the renormalization is necessary for the
point-like monopole parts only. Of course, we could expect this result as in
the region $r>a$ the local geometry is the same in both models and, hence,
the divergences are the same as well.

As it has been already mentioned, if there is no surface energy-momentum
tensor on the bounding surface $r=a$, then the radial derivatives of the
metric are continuous and, hence, in Eqs. (\ref{Fnutilde}) and (\ref%
{fetasbar}) one has $u_{a}^{\prime }=0$, $h_{a}^{\prime }=1/a$. In models
with an additional infinitely thin spherical shell located at $r=a$, these
quantities are related to the components of the corresponding surface
energy-momentum tensor $\tau _{\mu \nu }$. Denoting by $n^{\mu }$ the normal
to the shell normalized by the condition $n_{\mu }n^{\mu }=-1$ and assuming
that it points into the bulk on both sides, from the Israel matching
conditions one has%
\begin{equation}
\left\{ K_{\mu \nu }-Kh_{\mu \nu }\right\} =8\pi G\tau _{\mu \nu }.
\label{matchcond}
\end{equation}%
In this formula the curly brackets denote summation over each side of the
shell, $h_{\mu \nu }=g_{\mu \nu }+n_{\mu }n_{\nu }$ is the induced metric on
the shell, $K_{\mu \nu }=h_{\mu }^{\rho }h_{\nu }^{\delta }\nabla _{\rho
}n_{\delta }$ its extrinsic curvature and $K=K_{\mu }^{\mu }$. For the
region $r\leqslant a$ one has $n_{\mu }=\delta _{\mu }^{1}e^{v(r)}$ and the
non-zero components of the extrinsic curvature are given by the formulae%
\begin{equation}
K_{0}^{0}=-u^{\prime }(r)e^{-v(r)},\;K_{2}^{2}=K_{3}^{3}=-h^{\prime
}(r)e^{-v(r)},\;r=a-.  \label{exttensor}
\end{equation}%
The corresponding expressions for the region $r\geqslant a$ are obtained by
taking in these formulae $u(r)=v(r)=0$, $h(r)=\ln (\alpha r)$ and changing
the signs for the components of the extrinsic curvature tensor. Now from the
matching conditions (\ref{matchcond}) we find%
\begin{equation}
u_{a}^{\prime }=8\pi G\left( \tau _{2}^{2}-\frac{1}{2}\tau _{0}^{0}\right)
,\quad h_{a}^{\prime }=\frac{1}{a}+4\pi G\tau _{0}^{0}\ .  \label{matchcond2}
\end{equation}%
Note that the combination in the square brackets in Eqs. (\ref{Fnutilde})
and (\ref{fetasbar}) is related to the surface energy-momentum tensor by the
formula%
\begin{equation}
au_{a}^{\prime }/2+ah_{a}^{\prime }-1/2=2\pi Ga\tau +1/2\ .
\label{Tracesurf}
\end{equation}%
where $\tau $ is the trace of the surface energy-momentum tensor.

\section{Vacuum polarization in the flower-pot model}

\label{sec:flowerpot}

In this section, as an application of the general results given above we
consider a simple example of the core model assuming that the spacetime
inside it is flat. The corresponding model for the cosmic string core was
considered in Refs. \cite{Alle90,Alle96,Beze06b} and for the global monopole
core in Ref. \cite{S-E}. Following these papers we will refer to this model
as flower-pot model. Taking in the region inside the core $u(r)=v(r)=0$,
from the zero curvature condition one finds $e^{h(r)}=r+\mathrm{const}$. The
value of the constant is found from the continuity condition for the
function $h(r)$ at the boundary which gives $\mathrm{const}=(\alpha -1)a$.
Hence, in the flower-pot model the interior line element has the form%
\begin{equation}
ds^{2}=dt^{2}-dr^{2}-\left[ r+(\alpha -1)a\right] ^{2}(d\theta ^{2}+\sin
^{2}\theta d\phi ^{2}).  \label{intflow}
\end{equation}%
In terms of the radial coordinate $r$ the origin is located at $%
r=r_{0}=(1-\alpha )a$. From the matching conditions (\ref{matchcond2}) we
find the corresponding surface energy-momentum tensor with the non-zero
components and the trace given by%
\begin{equation}
\tau _{0}^{0}=2\tau _{2}^{2}=2\tau _{3}^{3}=\frac{1/\alpha -1}{4\pi Ga}%
,\;2\pi Ga\tau =\frac{1}{\alpha }-1.  \label{surfemtflow}
\end{equation}%
The corresponding surface energy density is positive for the global monopole
with $\alpha <1$. We will consider the VEVs in the exterior and interior
regions separately.

\subsection{Exterior region}

\label{subsec:flpotExt}

In the region inside the core the radial eigenfunctions regular at the
origin are the functions%
\begin{equation}
R_{1,n_{\sigma }}(r,k)=C_{\beta }\frac{J_{\lambda _{\sigma }}(k\tilde{r})}{%
\sqrt{\tilde{r}}},\;R_{2,n_{\sigma }}(r,k)=C_{\beta }\frac{J_{\lambda
_{\sigma }+n_{\sigma }}(k\tilde{r})}{\sqrt{\tilde{r}}}\ ,  \label{Rlflow}
\end{equation}%
where%
\begin{equation}
\lambda _{\sigma }=j+1/2-n_{\sigma }/2,  \label{lamdasigma}
\end{equation}%
and $\tilde{r}=r+(\alpha -1)a$ is the standard Minkowskian radial
coordinate, $0\leqslant \tilde{r}\leqslant \alpha a$. Note that for an
interior Minkowskian observer the radius of the core is $\alpha a$. The
normalization coefficient $C_{\beta }$ is found from the condition (\ref{R1a}%
):%
\begin{equation}
C_{\beta }^{2}=\frac{M+\omega }{\pi ^{2}\alpha \omega }\frac{2kJ_{\lambda
_{\sigma }}^{-2}(k\alpha a)}{\tilde{J}_{\nu _{\sigma }}^{2}(ka)+\tilde{Y}%
_{\nu _{\sigma }}^{2}(ka)},  \label{Clnorm}
\end{equation}%
with the tilted notation for the cylindrical functions%
\begin{equation}
\tilde{F}_{\nu _{\sigma }}(x)=xF_{\nu _{\sigma }}^{\prime }(x)-\left[ x\frac{%
J_{\lambda _{\sigma }}^{\prime }(\alpha x)}{J_{\lambda _{\sigma }}(\alpha x)}%
+\frac{1}{2}\left( \frac{1}{\alpha }-1\right) \right] F_{\nu _{\sigma }}(x).
\label{FtildeFlow}
\end{equation}%
Note that $\tilde{J}_{\nu _{\sigma }}(x)=0$ for $\alpha =1$. In this case
for the barred notation in Eq. (\ref{fetasbar}) one has $\bar{f}^{(-1,0)}(x)=
$ $\bar{f}^{(1,0)}(x)$. Hence, in the flower-pot model the part in the
energy-momentum tensor due the non-trivial structure of the core is given by
the formula
\begin{eqnarray}
\langle T_{\mu }^{\nu }\rangle _{\mathrm{c}} &=&\frac{\delta _{\mu }^{\nu }}{%
\pi ^{2}\alpha ^{2}r}\sum_{l=1}^{\infty }l\int_{M}^{\infty }dx\frac{x^{3}}{%
\sqrt{x^{2}-M^{2}}}  \notag \\
&&\times \frac{C\left\{ I_{l-1/2}(\alpha ax),I_{l/\alpha -1/2}(ax)\right\} }{%
C\left\{ I_{l-1/2}(\alpha ax),K_{l/\alpha -1/2}(ax)\right\} }G_{l/\alpha
-1/2}^{(\mu )}\left[ x,K_{l/\alpha -1/2}(xr)\right] ,  \label{Tmunu8}
\end{eqnarray}%
where we have introduced the notation%
\begin{equation}
C\left\{ f(\alpha x),g(x)\right\} =xf(\alpha x)g^{\prime }(x)-\left[ \frac{1%
}{2}\left( \frac{1}{\alpha }-1\right) f(\alpha x)+xf^{\prime }(\alpha x)%
\right] g(x).  \label{Cfg}
\end{equation}%
In formula (\ref{Tmunu8}), $G_{\nu }^{(\mu )}\left[ x,f_{\nu }(y)\right]
=F_{\nu }^{(\mu ,\eta _{s})}\left[ x,f_{\nu }(y)\right] $ for $\mu =1,2,3$,
and
\begin{equation}
G_{l/\alpha -1/2}^{(0)}\left[ x,f_{l/\alpha -1/2}(y)\right] =\left( \frac{%
M^{2}}{x^{2}}-1\right) \left[ f_{l/\alpha -1/2}^{2}(y)-f_{l/\alpha
+1/2}^{2}(y)\right] .  \label{Gnu0}
\end{equation}%
Note that in terms of notation (\ref{Cfg}) one has
\begin{equation}
J_{\lambda _{\sigma }}(\alpha x)\tilde{F}_{\nu _{\sigma }}(x)=C\{J_{\lambda
_{\sigma }}(\alpha x),F_{\nu _{\sigma }}(x)\}.  \label{bartoC}
\end{equation}%
For $\alpha =1$ we have $C\left\{ I_{l-1/2}(\alpha ax),I_{l/\alpha
-1/2}(ax)\right\} =0$ and the VEVs vanish. In the similar way, for the
fermionic condensate from Eq. (\ref{condens1}) we find%
\begin{eqnarray}
\langle \bar{\psi}\psi \rangle _{\mathrm{c}} &=&\frac{M}{\pi ^{2}\alpha ^{2}r%
}\sum_{l=1}^{\infty }l\int_{M}^{\infty }dx\,\frac{x}{\sqrt{x^{2}-M^{2}}}
\notag \\
&&\times \frac{C\left\{ I_{l-1/2}(\alpha ax),I_{l/\alpha -1/2}(ax)\right\} }{%
C\left\{ I_{l-1/2}(\alpha ax),K_{l/\alpha -1/2}(ax)\right\} }\left[
K_{l/\alpha -1/2}^{2}(xr)-K_{l/\alpha +1/2}^{2}(xr)\right] .  \label{condfp}
\end{eqnarray}%
For a massless fermionic field the core-induced part in the condensate
vanishes. From formulae (\ref{Tmunu8}) and (\ref{condfp}) it can be seen
that for fixed values of $r$ and $M$, in the limit $a\rightarrow 0$ the
core-induced parts vanish as $a^{2/\alpha +1}$. Note that by using the
recurrence relations for the modified Bessel functions, the functions $%
C\left\{ f(\alpha y),g(y)\right\} $ in these formulae can also be written in
the form%
\begin{equation}
C\left\{ I_{l-1/2}(\alpha y),f_{l/\alpha -1/2}(y)\right\} =\lambda
_{f}yI_{l-1/2}(\alpha y)f_{l/\alpha +1/2}(y)-yI_{l+1/2}(\alpha y)f_{l/\alpha
-1/2}(y),  \label{CIf}
\end{equation}%
with $f=I$, $K$ and $\lambda _{I}=1$, $\lambda _{K}=-1$. In particular, from
formula (\ref{CIf}) with $f=K$ it follows that $C\left\{ I_{l-1/2}(\alpha
y),K_{l/\alpha -1/2}(y)\right\} <0$. As we will show in Appendix \ref%
{sec:app1}, this means that there are no bound states in the flower-pot
model.

The core-induced part in the VEVs of the energy-momentum tensor and the
fermionic condensate given by formulae (\ref{Tmunu8}) and (\ref{condfp}),
are finite for $r>a$ and diverge on the core boundary. Surface divergences
are well-known in quantum field theory with boundaries and are investigated
for various boundary geometries and fields. To find the corresponding
asymptotic behavior for the points near the boundary, we note that in this
region the main contribution comes from large values of $l$. By using the
uniform asymptotic expansions for the modified Bessel functions for large
values of the order (see, for instance, \cite{Abra64}), to the leading order
we find%
\begin{eqnarray}
\langle T_{0}^{0}\rangle _{\mathrm{c}} &\approx &-2\langle T_{2}^{2}\rangle
_{\mathrm{c}}\approx \frac{2a}{r-a}\langle T_{1}^{1}\rangle _{\mathrm{c}%
}\approx \frac{\alpha -1}{120\pi ^{2}\alpha a(r-a)^{3}},\;  \label{nearcore}
\\
\langle \bar{\psi}\psi \rangle _{\mathrm{c}} &\approx &\frac{(1-\alpha )M}{%
12\pi ^{2}\alpha a(r-a)}.  \label{nearcorecond}
\end{eqnarray}%
As the parts corresponding to the geometry of the point-like global monopole
are finite at $r=a$, from here we conclude that near the core the VEVs are
dominated by the core-induced parts.

At large distances from the core, $r\gg a$, in the case of a massless
fermionic field we introduce a new integration variable $y=xr$ and expand
the integrand over $a/r$. The main contribution comes from the lowest mode $%
l=1$ and to the leading order one has%
\begin{equation}
\langle T_{\mu }^{\nu }\rangle _{\mathrm{c}}\approx -\frac{\delta _{\mu
}^{\nu }}{3\pi ^{2}\alpha a^{4}}\frac{(1-\alpha )2^{1-2/\alpha
}(a/r)^{2/\alpha +5}}{(2+\alpha )\Gamma ^{2}(1/\alpha +1/2)}\int_{0}^{\infty
}dy\,y^{2/\alpha +3}G_{1/\alpha -1/2}^{(\mu )}\left[ y/r,K_{1/\alpha -1/2}(y)%
\right] .  \label{TmunuLargedist}
\end{equation}%
Note that for a massless field the integrand does not depend on $r$. The
integrals in Eq. (\ref{TmunuLargedist}) may be evaluated by using the
formula for the integrals involving the product of the MacDonald functions
(see Ref. \cite{Prud86}). For a massive field, assuming $Mr\gg 1$, the main
contribution into the integral over $x$ comes from the lower limit of the
integration. Replacing the functions $K_{l/\alpha \pm 1/2}(xr)$ by the
corresponding asymptotic formulae for large values of the argument, to the
leading order we obtain the following estimates%
\begin{equation}
\langle T_{0}^{0}\rangle _{\mathrm{c}}\approx -\langle T_{1}^{1}\rangle _{%
\mathrm{c}}\approx \frac{1}{Mr}\langle T_{2}^{2}\rangle _{\mathrm{c}}\approx
-\frac{\exp (-2Mr)}{4\sqrt{\pi Mr}\alpha ^{3}r^{4}}\sum_{l=1}^{\infty }l^{2}%
\frac{C\left\{ I_{l-1/2}(\alpha aM),I_{l/\alpha -1/2}(aM)\right\} }{C\left\{
I_{l-1/2}(\alpha aM),K_{l/\alpha -1/2}(aM)\right\} }.  \label{largemassiv}
\end{equation}%
As we see, in this limit the core-induced energy density and the radial
stress are suppressed by the factor $Mr$ with respect to the azimuthal
stress.

For $\alpha \ll 1$ the solid angle for the exterior geometry is small and
the corresponding scalar curvature is large. In this limit we replace the
modified Bessel function containing in the index $l/\alpha $ by the
corresponding uniform asymptotic expansions for large values of the order
and the functions containing in the argument $\alpha ax$ by the expansions
for small values of the argument. In this way it can be seen that the main
contribution into the VEVs comes from the mode $l=1$ and the VEVs are
suppressed by the factor $\exp [-(2/\alpha )\ln (r/a)]$.

\begin{figure}[tbph]
\begin{center}
\epsfig{figure=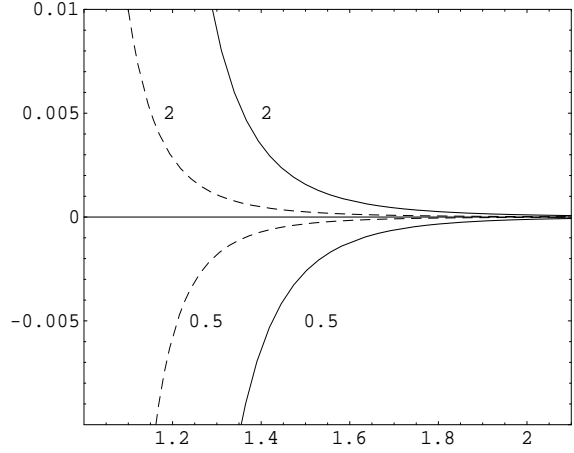,width=7.5cm,height=6cm}
\end{center}
\caption{The core-induced energy density (full curves),
$a^{4}\langle T_{0}^{0}\rangle _{\mathrm{c}}$, and radial stress
(dashed curves), $a^{4}\langle T_{1}^{1}\rangle _{\mathrm{c}}$,
for a massless fermionic field in the exterior region of the
flower-pot model as functions on $r/a$. The numbers near the
curves correspond to the values of the parameter $\protect\alpha
$.} \label{fig1}
\end{figure}

\begin{figure}[tbph]
\begin{center}
\epsfig{figure=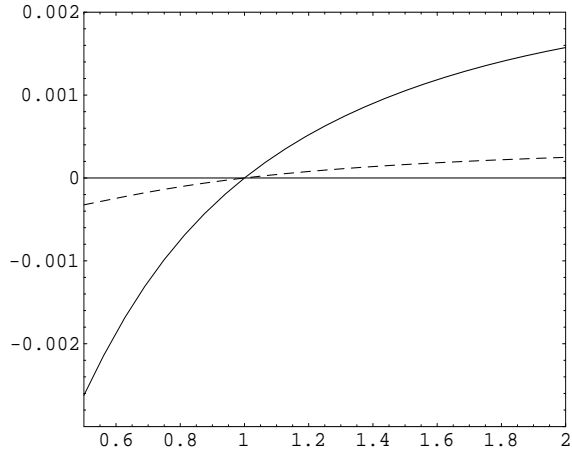,width=7.5cm,height=6cm}
\end{center}
\caption{The core-induced energy density (full curves),
$a^{4}\langle T_{0}^{0}\rangle _{\mathrm{c}}$, and radial stress
(dashed curves), $a^{4}\langle T_{1}^{1}\rangle _{\mathrm{c}}$,
for a massless fermionic field evaluated at $r/a=1.5$ as functions
on the parameter $\protect\alpha $.} \label{fig2}
\end{figure}

In figure \ref{fig1} we have plotted the dependence of the core-induced
energy density (full curves) and radial stress (dashed curves) for a
massless fermionic field as functions on the scaled radial coordinate $r/a$
for $\alpha =0.5$ and $\alpha =2$. The azimuthal stresses are found from the
zero trace condition. In figure \ref{fig2} the same quantities evaluated for
$r/a=1.5$ are presented as functions on the parameter $\alpha $.

\subsection{Interior region}

\label{subsec:flpotinter}

Now let us consider the vacuum polarization effects inside the core for the
flower-pot model. The corresponding eigenfunctions have the form given by
Eq. (\ref{psibetin}) where the functions $R_{1,n_{\sigma }}(r,k)$ and $%
R_{2,n_{\sigma }}(r,k)$ are defined by formulae (\ref{Rlflow}). Substituting
the eigenfunctions into the mode-sum formula, for the corresponding
energy-momentum tensor one finds (no summation over $\mu $)%
\begin{equation}
\langle 0|T_{\mu }^{\nu }|0\rangle =\frac{\delta _{\mu }^{\nu }}{2\pi
^{3}\alpha a^{3}\tilde{r}}\sum_{j=1/2}^{\infty }(2j+1)\sum_{\sigma
=0,1}\int_{0}^{\infty }dx\frac{f_{\sigma \lambda _{\sigma }}^{(\mu )}\left[
x,J_{\lambda _{\sigma }}(x\tilde{r}/a)\right] }{J_{\lambda _{\sigma
}}^{2}(\alpha x)\left[ \tilde{J}_{\nu _{\sigma }}^{2}(x)+\tilde{Y}_{\nu
_{\sigma }}^{2}(x)\right] }.  \label{Tmunuinfl}
\end{equation}%
To find the renormalized VEV of the energy-momentum tensor we need to
evaluate the difference between the VEV given by Eq. (\ref{Tmunuinfl}) and
the corresponding VEV for the Minkowski bulk:%
\begin{equation}
\langle T_{\mu }^{\nu }\rangle _{\mathrm{ren}}=\langle 0|T_{\mu }^{\nu
}|0\rangle -\langle 0_{M}|T_{\mu }^{\nu }|0_{M}\rangle .  \label{Tmunusubin}
\end{equation}%
The appropriate form for the Minkowskian part is obtained from Eq. (\ref%
{qext}) taking $\alpha =1$, replacing the integrand by $f_{\sigma \lambda
_{\sigma }}^{(\mu )}\left[ x,J_{\lambda _{\sigma }}(x\tilde{r}/a)\right] $
and $r\rightarrow \tilde{r}$. As a result for the subtracted VEV one finds%
\begin{eqnarray}
\langle T_{\mu }^{\nu }\rangle _{\mathrm{ren}} &=&\frac{\delta _{\mu }^{\nu }%
}{2\pi ^{3}\alpha a^{3}\tilde{r}}\sum_{j=1/2}^{\infty }(2j+1)\sum_{\sigma
=0,1}\int_{0}^{\infty }dxf_{\sigma \lambda _{\sigma }}^{(\mu )}\left[
x,J_{\lambda _{\sigma }}(x\tilde{r}/a)\right]   \notag \\
&&\times \left[ \frac{J_{\lambda _{\sigma }}^{-2}(\alpha x)/\alpha }{\tilde{J%
}_{\nu _{\sigma }}^{2}(x)+\tilde{Y}_{\nu _{\sigma }}^{2}(x)}-\frac{\pi ^{2}}{%
4}\right] .  \label{Tmunusub2}
\end{eqnarray}%
The integral in this formula is slowly convergent and the integrand is
highly oscillatory.

In order to transform the expression for the subtracted VEV of the
energy-momentum tensor into more convenient form, we note that the following
identity takes place%
\begin{equation}
\frac{J_{\lambda _{\sigma }}^{-2}(\alpha x)/\alpha }{\tilde{J}_{\nu _{\sigma
}}^{2}(x)+\tilde{Y}_{\nu _{\sigma }}^{2}(x)}-\frac{\pi ^{2}}{4}=-\frac{\pi
^{2}}{8}\sum_{s=1,2}\frac{C\{H_{\lambda _{\sigma }}^{(s)}(\alpha x),H_{\nu
_{\sigma }}^{(s)}(x)\}}{C\{J_{\lambda _{\sigma }}(\alpha x),H_{\nu _{\sigma
}}^{(s)}(x)\}},  \label{identflow2new}
\end{equation}%
Substituting Eq. (\ref{identflow2new}) into formula (\ref{Tmunusub2}), we
rotate the integration contour in the complex plane $x$ by the angle $\pi /2$
for $s=1$ and by the angle $-\pi /2$ for $s=2$. Under the condition $\tilde{r%
}<\alpha a$ the contributions from the semicircles with the radius tending
to infinity vanish. The integrals over the segments $(0,iMa)$ and $(0,-iMa)$
of the imaginary axis cancel out and after introducing the modified Bessel
functions we can see that different parities give the same contribution.
Consequently, the renormalized VEV for the energy-momentum tensor can be
presented in the form (no summation over $\mu $)%
\begin{eqnarray}
\langle T_{\mu }^{\nu }\rangle _{\mathrm{ren}} &=&\frac{\delta _{\mu }^{\nu }%
}{\pi ^{2}\alpha \tilde{r}}\sum_{l=1}^{\infty }l\int_{M}^{\infty }dx\frac{%
x^{3}}{\sqrt{x^{2}-M^{2}}}  \notag \\
&&\times \frac{C\{K_{l-1/2}(\alpha ax),K_{l/\alpha -1/2}(ax)\}}{%
C\{I_{l-1/2}(\alpha ax),K_{l/\alpha -1/2}(ax)\}}G_{l-1/2}^{(\mu )}\left[
x,I_{l-1/2}(x\tilde{r})\right] ,  \label{Tmunusub3}
\end{eqnarray}%
where the functions $G_{\nu }^{(\mu )}\left[ x,f_{\nu }(y)\right] $ are the
same as in Eq. (\ref{Tmunu8}) with $\lambda _{I}=1$ in Eqs. (\ref{Fnu1}) and
(\ref{Fnu2}). Note that similar to Eq. (\ref{CIf}), the function in the
numerator of the integrand is also presented in the form%
\begin{equation}
C\left\{ K_{l-1/2}(\alpha y),K_{l/\alpha -1/2}(y)\right\}
=-yK_{l-1/2}(\alpha y)K_{l/\alpha +1/2}(y)+yK_{l+1/2}(\alpha y)K_{l/\alpha
-1/2}(y).  \label{CKK}
\end{equation}%
For $\alpha =1$ one has $C\{K_{l-1/2}(\alpha ax),K_{l/\alpha -1/2}(ax)\}=0$
and, as we could expect, the VEV of the energy-momentum tensor vanishes. It
can be explicitly checked that the components of the tensor given by formula
(\ref{Tmunusub3}) satisfy the continuity equation (\ref{conteq}) and this
tensor is traceless for a massless field. In the way similar to that for the
exterior region, for the renormalized fermionic condensate inside the core
we find%
\begin{eqnarray}
\langle \bar{\psi}\psi \rangle _{\mathrm{ren}} &=&\frac{M}{\pi ^{2}\alpha
\tilde{r}}\sum_{l=1}^{\infty }l\int_{M}^{\infty }dx\,\frac{x}{\sqrt{%
x^{2}-M^{2}}}  \notag \\
&&\times \frac{C\{K_{l-1/2}(\alpha ax),K_{l/\alpha -1/2}(ax)\}}{%
C\{I_{l-1/2}(\alpha ax),K_{l/\alpha -1/2}(ax)\}}\left[ I_{l-1/2}^{2}(x\tilde{%
r})-I_{l+1/2}^{2}(x\tilde{r})\right] .  \label{condInt}
\end{eqnarray}

The VEVs given by formulae (\ref{Tmunusub3}) and (\ref{condInt}) are finite
for $\tilde{r}<\alpha a$ and diverge on the core boundary. For the points
near the boundary the main contribution comes from large $l$ and we replace
the modified Bessel functions by the corresponding uniform expansions for
large values of the order. In this way it can be seen that the leading terms
in the asymptotic expansion with respect to the distance from the boundary
are given by the formulae%
\begin{eqnarray}
\langle T_{0}^{0}\rangle _{\mathrm{ren}} &\approx &-2\langle
T_{2}^{2}\rangle _{\mathrm{ren}}\approx -\frac{2\alpha a}{\alpha a-\tilde{r}}%
\langle T_{1}^{1}\rangle _{\mathrm{ren}}\approx \frac{\alpha -1}{120\pi
^{2}\alpha ^{2}a(\alpha a-\tilde{r})^{3}},  \label{nearInt} \\
\langle \bar{\psi}\psi \rangle _{\mathrm{ren}} &\approx &\frac{(1-\alpha )M}{%
12\pi ^{2}\alpha ^{2}a(\alpha a-\tilde{r})}.  \label{nearboundInt}
\end{eqnarray}%
Comparing these results with the corresponding formulae for the exterior
region, we see that near the core boundary the energy density and azimuthal
stresses have the same signs inside and outside the core, whereas the radial
stresses have opposite signs.

Near the core center the contribution of the summand with a given $l$
behaves like $\tilde{r}^{2(l-1)}$ and the main contribution comes from the
lowest mode $l=1$ with the leading term (no summation over $\mu $)%
\begin{equation}
\langle T_{\mu }^{\nu }\rangle _{\mathrm{ren}}\approx \frac{2\delta _{\mu
}^{\nu }}{3\pi ^{3}\alpha a^{4}}\int_{Ma}^{\infty }dx\frac{x^{3}C^{(\mu )}(x)%
}{\sqrt{x^{2}-M^{2}a^{2}}}\frac{C\{K_{1/2}(\alpha x),K_{1/\alpha -1/2}(x)\}}{%
C\{I_{1/2}(\alpha x),K_{1/\alpha -1/2}(x)\}},  \label{nearcenter}
\end{equation}%
where we have introduced notations%
\begin{equation}
C^{(0)}(x)=3(M^{2}a^{2}/x^{2}-1),\;C^{(1)}(x)=C^{(2)}(x)=1.  \label{Cmux}
\end{equation}%
In the case of a massless field, formula (\ref{nearcenter}) also gives the
behavior of the vacuum energy-momentum tensor in the limit when the core
radius is large and $\tilde{r}$ is fixed, $a/\tilde{r}\gg 1$. In the same
limit, for a massive field, assuming $aM\gg 1$, we replace in Eq. (\ref%
{Tmunusub3}) \ the modified Bessel functions containing in the argument $ax$
by the corresponding asymptotic expressions for large values of the
argument. By taking into account that the main contribution into the
integral comes from the lower limit of the integration, to the leading order
we have (no summation over $\mu $)%
\begin{eqnarray}
\langle T_{0}^{0}\rangle _{\mathrm{ren}} &\approx &\frac{\alpha -1}{16\alpha
^{4}a^{3}\tilde{r}}\frac{e^{-2\alpha aM}}{\sqrt{\pi \alpha aM}}%
\sum_{l=1}^{\infty }l^{2}\left[ I_{l-1/2}^{2}(M\tilde{r})-I_{l+1/2}^{2}(M%
\tilde{r})\right] ,  \label{T00radlarge} \\
\langle T_{\mu }^{\nu }\rangle _{\mathrm{ren}} &\approx &\frac{\delta _{\mu
}^{\nu }(1-\alpha )}{8\pi \alpha ^{4}a^{3}\tilde{r}}\sqrt{\pi \alpha aM}%
e^{-2\alpha aM}\sum_{l=1}^{\infty }l^{2}G_{l-1/2}^{(\mu )}\left[
M,I_{l-1/2}(M\tilde{r})\right] ,  \label{Tmnradlarge}
\end{eqnarray}%
with $\;\nu \neq 0$. In this case the energy density is suppressed with
respect to the vacuum stresses by an additional factor $(aM)^{-1}$.

For small values of the parameter $\alpha $ assuming that the core radius $%
\alpha a$ for an internal Minkowskian observer is fixed, we replace the
functions $K_{1/\alpha -1/2}(ax)$ in Eq. (\ref{Tmunusub3}) by the
corresponding uniform asymptotic expansion for large values of the order.
The leading term is obtained by making use of the replacements%
\begin{equation}
C\{f_{l-1/2}(\alpha ax),K_{1/\alpha -1/2}(ax)\}\rightarrow y[\ln
f_{l-1/2}(y)]^{\prime }+\sqrt{l^{2}+y^{2}}+1/2,  \label{replace}
\end{equation}
in the integrand of Eq. (\ref{Tmunusub3}) with $y=\alpha ax$ and $f=I,K$. As
a result, in this limit the VEVs behave like $1/\alpha $. Due to the factor $%
\alpha ^{2}$ in the volume element the corresponding global quantities, such
as total energy vanish as $\alpha $.

\begin{figure}[tbph]
\begin{center}
\epsfig{figure=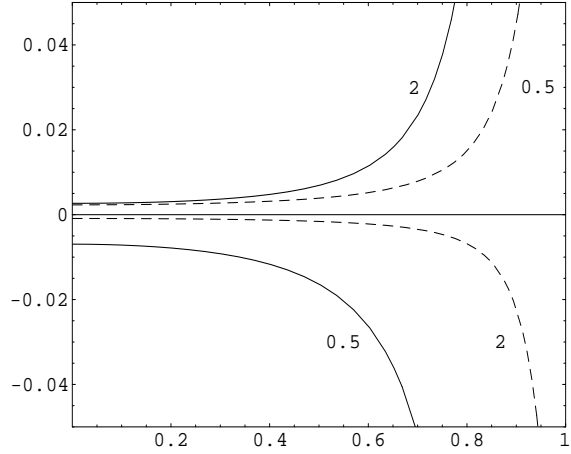,width=7.5cm,height=6cm}
\end{center}
\caption{The renormalized vacuum energy density (full curves), $(\protect%
\alpha a)^{4}\langle T_{0}^{0}\rangle _{\mathrm{sub}}$, and radial stress
(dashed curves), $(\protect\alpha a)^{4}\langle T_{1}^{1}\rangle _{\mathrm{%
sub}}$, for a massless fermionic field inside the core of the flower-pot
model as functions on $\tilde{r}/\protect\alpha a$. The numbers near the
curves correspond to the values of the parameter $\protect\alpha $.}
\label{fig3}
\end{figure}

\begin{figure}[tbph]
\begin{center}
\epsfig{figure=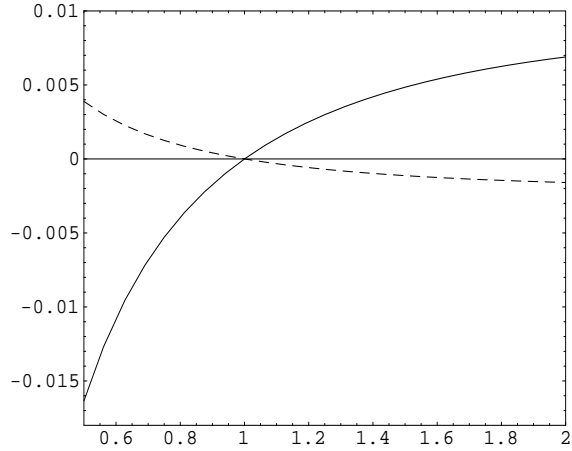,width=7.5cm,height=6cm}
\end{center}
\caption{The energy density (full curves), $(\protect\alpha a)^{4}\langle
T_{0}^{0}\rangle _{\mathrm{sub}}$, and radial stress (dashed curves), $(%
\protect\alpha a)^{4}\langle T_{1}^{1}\rangle _{\mathrm{sub}}$, for a
massless fermionic field evaluated at $\tilde{r}/\protect\alpha a=0.5$ as
functions on the parameter $\protect\alpha $.}
\label{fig4}
\end{figure}

In figure \ref{fig3} we have plotted the dependence of the renormalized
vacuum energy density (full curves) and radial stress (dashed curves) for a
massless fermionic field as functions on $\tilde{r}/\alpha a$ for $\alpha
=0.5$ and $\alpha =2$. In figure \ref{fig4} the same quantities evaluated
for $\tilde{r}/\alpha a=0.5$ are presented as functions on the parameter $%
\alpha $.

\section{Conclusion}

\label{sec:Conc}

In the present paper we have considered the polarization of the fermionic
vacuum by the gravitational field of the global monopole wit a non-trivial
core structure. The previous investigations in this direction are concerned
with the idealized point-like model, where the curvature has singularity at
the origin. In a realistic point of view, the global monopole has a
characteristic core radius determined by the symmetry braking energy scale.
For a general spherically symmetric static model of the core with finite
thickness, we have evaluated the VEVs of the energy-momentum tensor and the
fermionic condensate for a massive spinor field. These quantities are among
the most important characteristics of the vacuum properties, which carry an
information about the core structure. In the region outside the core we have
presented the VEVs as a sum of two contributions. The first one corresponds
to the geometry of a point-like global monopole and the second one is
induced by the non-trivial structure of the monopole core. In the general
spherically symmetric static model for the core, we have derived closed
analytic expressions for the core-induced parts given by formula (\ref%
{Tmunu6}) for the energy-momentum tensor and by formula (\ref{condens1}) for
the fermionic condensate, where the properties of the core are codified by
the coefficient in the square brackets in the notation (\ref{fetasbar}). In
Appendix \ref{sec:app1} we show that the formulae for the core-induced parts
in the VEVs of the energy-momentum tensor and fermionic condensate are valid
also in the case when bound states are present. For points away from the
core boundary these parts are finite and the renormalization is reduced to
that for the point-like monopole geometry. Of course, we could expect this
result as in the model under consideration the exterior geometry is the same
as that for the point-like global monopole. For the points on the boundary
the VEVs contain surface divergences well-known in quantum field theory with
boundaries.

As an example of the application of the general results, in section \ref%
{sec:flowerpot} we have considered a simple model with a flat spacetime
inside the core. The corresponding model for the cosmic string core is known
in literature as flower-pot model and here we use the same terminology for
the global monopole. To have matching between the exterior and interior
metrics, in this model we need the surface energy-momentum tensor located on
the boundary of the core and having components given by Eq. (\ref%
{surfemtflow}). In the model with solid angle deficit ($\alpha <1$) the
corresponding surface energy density is positive. The core-induced parts of
the exterior VEVs in the flower-pot model are obtained from the general
results taking as the interior radial functions in the eigenmodes the
functions (\ref{Rlflow}). These parts are given by formula (\ref{Tmunu8})
for the energy-momentum tensor and by formula (\ref{condfp}) for the
fermionic condensate. We have investigated the core-induced parts in various
asymptotic regions of the parameters. In the limit when the core radius
tends to zero, $a\rightarrow 0$, for fixed values $r$ and $M$, these parts
behave like $a^{2/\alpha +1}$. For points near the core boundary the leading
terms in the corresponding asymptotic expansions are given by formulae (\ref%
{nearcore}), (\ref{nearcorecond}). In this region the total VEVs are
dominated by the core-induced parts. At large distances from the core these
parts tend to zero as $(a/r)^{2/\alpha +5}$ for a massless field and are
exponentially suppressed by the factor $\exp (-Mr)$ for a massive field. We
have also investigated the limit of strong gravitational fields
corresponding to small values of the parameter $\alpha $. In this limit the
main contribution into the VEVs comes from the lowest mode $l=1$ and the
VEVs are suppressed by the factor $\exp [-(2/\alpha )\ln (r/a)]$.

For the flower-pot model we have also investigated the VEVs of the
energy-momentum tensor and the fermionic condensate inside the core. Though
the corresponding spacetime geometry is Monkowskian, the non-trivial
topology of the exterior region induces vacuum polarization effects in this
region as well. The renormalization is achieved by the subtraction from the
mode-sums the corresponding \ quantities for the Minkowski spacetime. By
making use of identity (\ref{identflow2new}), after an appropriate
deformation of the integration contour, we have presented the renormalized
VEV of the energy-momentum tensor in the form (\ref{Tmunusub3}) and the
fermionic condensate in the form (\ref{condInt}). These quantities are
finite for strictly interior points and diverge on the boundary of the core
with the leading divergences given by formulae (\ref{nearInt}), (\ref%
{nearboundInt}). In particular, near the core boundary the energy density
and azimuthal stresses have the same signs inside and outside the core,
whereas the radial stresses have opposite signs. Near the core center the
main contribution comes from the mode $l=1$ and the VEVs tend to a finite
limiting value with isotropic vacuum stresses. Although the exact behavior
for the fermionic field is unknown for a realistic model of the global
monopole spacetime, the flower-pot model considered here presents some
expected results as, for example, finite vacuum polarization effects at the
monopole's center. For large values of the core radius the renormalized VEV
of the energy-momentum tensor inside the core vanishes as $a^{-4}$ for a
massless field and as $e^{-2\alpha aM}$ for a massive one. In the limit $%
\alpha \ll 1$, assuming that the core radius $\alpha a$ for an internal
Minkowskian observer is fixed, the vacuum densities in the interior region
behave as $1/\alpha $.

Note that in this paper we have considered quantum vacuum effects in a
prescribed background, i.e. the gravitational back-reaction of quantum
effects is not taken into account. This back-reaction could have important
effects on the dynamical evolution of the bulk model. We do not consider
this extension of the theory, but note that the results presented here
constitute the starting point for such investigations.

\section*{Acknowledgement}

AAS was supported by PVE/CAPES Program and in part by the Armenian Ministry
of Education and Science Grant No. 0124. ERBM thanks Conselho Nacional de
Desenvolvimento Cient\'{\i}fico e Tecnol\'{o}gico (CNPq) and FAPESQ-PB/CNPq
(PRONEX) for partial financial support.

\appendix

\section{Bound states}

\label{sec:app1}

In this appendix we consider the changes in the procedure described in the
main text when bound states are present. For these states the quantity $k$
is purely imaginary, $k=i\gamma $, and the corresponding exterior
eigenfunctions have the form%
\begin{equation}
\psi _{\mathrm{b}\beta }^{\mathrm{(ex)}}=C_{\mathrm{b}}\frac{e^{-i\omega t}}{%
\sqrt{r}}\left(
\begin{array}{c}
K_{\nu _{\sigma }}(\gamma r)\Omega _{jl_{\sigma }m}(\theta ,\varphi ) \\
\frac{i\gamma }{\omega +M}K_{\nu _{\sigma }+n_{\sigma }}(\gamma r)(\hat{n}%
\cdot \vec{\sigma})\Omega _{jl_{\sigma }m}(\theta ,\varphi )%
\end{array}%
\right) ,  \label{psibound}
\end{equation}%
with $\omega ^{2}=M^{2}-\gamma ^{2}$. To have a stable ground state we will
assume that $\gamma <M$. From the continuity of the eigenfunctions at $r=a$
one has%
\begin{equation}
\sqrt{a}R_{1,n_{\sigma }}(a,i\gamma )=C_{\mathrm{b}}K_{\nu _{\sigma
}}(\gamma a),\;\sqrt{a}R_{2,n_{\sigma }}(a,i\gamma )=-in_{\sigma }C_{\mathrm{%
b}}K_{\nu _{\sigma }+n_{\sigma }}(\gamma a).  \label{appcond1}
\end{equation}%
Excluding from these relations the normalization coefficient $C_{\mathrm{b}}$
we see that for possible bound states $\gamma $ is a solutions of the
equation%
\begin{equation}
\bar{K}_{\nu _{\sigma }}^{(1,\sigma )}(\gamma a)=0,  \label{appBSeq}
\end{equation}%
with the barred notation from Eq. (\ref{fetasbar}).

The coefficient $C_{\mathrm{b}}$ in Eq. (\ref{psibound}) is found from the
normalization condition (\ref{normcond}). To derive a formula for the
normalization integral, we rewrite equations (\ref{f1}), (\ref{g1}) in terms
of the functions $F_{\beta }(r)=e^{u/2+h}f_{\beta }(r)$ and $G_{\beta
}(r)=e^{u/2+h}g_{\beta }(r)$ and differentiate both equations with respect
to $\omega $. Further we multiply the first equation by $G_{\beta }(r)$, the
second one by $-F_{\beta }(r)$ and add them. Combining the resulting
equation with Eqs. (\ref{f1}), (\ref{g1}), it can be seen that the following
relation takes place%
\begin{equation}
e^{v-u}\left( F_{\beta }^{2}+G_{\beta }^{2}\right) =e^{u+2h}\frac{d}{dr}%
\left( G_{\beta }\frac{\partial F_{\beta }}{\partial \omega }-F_{\beta }%
\frac{\partial G_{\beta }}{\partial \omega }\right) .  \label{relNorm}
\end{equation}%
Integrating this relation we obtain the formula%
\begin{equation}
\int_{r_{0}}^{r}dr\,e^{v+2h}[f_{\beta }^{2}(r)+g_{\beta }^{2}(r)]=e^{u+2h}
\left[ g_{\beta }(r)\frac{\partial f_{\beta }(r)}{\partial \omega }-f_{\beta
}(r)\frac{\partial g_{\beta }(r)}{\partial \omega }\right] ,  \label{normint}
\end{equation}%
where $r_{0}$ is the value of the radial coordinate corresponding to the
center of the core. By using this formula, in the case of bound states for
the normalization integral in Eq. (\ref{normcond}) one finds%
\begin{equation}
\int d^{3}x\sqrt{\gamma }\psi _{\beta }^{+}\psi _{\beta
}=\int_{r_{0}}^{\infty }dr\,e^{v+2h}[f_{\beta }^{2}(r)+g_{\beta
}^{2}(r)]=e^{u+2h}\left[ g_{\beta }(r)\frac{\partial f_{\beta }(r)}{\partial
\omega }-f_{\beta }(r)\frac{\partial g_{\beta }(r)}{\partial \omega }\right]
_{r=a+}^{r=a-}.  \label{normInt}
\end{equation}%
The expression on the right can be further simplified by using the
continuity of the eigenfunctions on the boundary of the core. In this way
for the normalization coefficient one finds the formula%
\begin{equation}
C_{\mathrm{b}}^{2}=\frac{M+\omega }{\alpha ^{2}K_{\nu _{\sigma }}(\gamma
a)(\partial /\partial \omega )\bar{K}_{\nu _{\sigma }}^{(1,\sigma )}(\gamma
a)}.  \label{Cb2}
\end{equation}

As a result, for the contribution of the bound state with $k=i\gamma $ to
the VEV\ of the energy-momentum tensor we have the formula (no summation
over $\mu $)%
\begin{equation}
\langle T_{\mu }^{\nu }\rangle _{\mathrm{b}}=\frac{\delta _{\mu }^{\nu }}{%
2\pi \alpha ^{2}ar}\sum_{l=1}^{\infty }l\sum_{\sigma =0,1}\frac{\gamma ^{3}}{%
\sqrt{M^{2}-\gamma ^{2}}}\frac{\bar{I}_{\nu _{\sigma }}^{(1,\sigma )}(\gamma
a)}{\bar{K}_{\nu _{\sigma }}^{(1,\sigma )\prime }(\gamma a)}B^{(\mu
)}[\gamma ,K_{\nu _{\sigma }}(\gamma r)],  \label{TmunuBound}
\end{equation}%
where $\nu _{\sigma }=l/\alpha -n_{\sigma }/2$ and we have introduced the
notations%
\begin{eqnarray}
B^{(0)}[\gamma ,K_{\nu _{\sigma }}(y)] &=&\left( 1-\frac{M^{2}}{\gamma ^{2}}%
\right) \left[ \left( \frac{M}{\sqrt{M^{2}-\gamma ^{2}}}-1\right)
K_{\nu _{\sigma }}^{2}(y) \right. \nonumber \\
&& \left. +\left( \frac{M}{\sqrt{M^{2}-\gamma ^{2}}}+1\right)
K_{\nu
_{\sigma }+n_{\sigma }}^{2}(y)\right] ,  \nonumber \\
B^{(1)}[\gamma ,K_{\nu _{\sigma }}(y)] &=&K_{\nu _{\sigma }}^{2}(y)+\frac{%
2ln_{\sigma }}{\alpha y}K_{\nu _{\sigma }}(y)K_{\nu _{\sigma }+n_{\sigma
}}(y)-K_{\nu _{\sigma }+n_{\sigma }}^{2}(y),  \label{Bmu} \\
B^{(\mu )}[\gamma ,K_{\nu _{\sigma }}(y)] &=&-\frac{ln_{\sigma }}{\alpha y}%
K_{\nu _{\sigma }}(y)K_{\nu _{\sigma }+n_{\sigma }}(y),\;\mu =2,3.
\nonumber
\end{eqnarray}%
In deriving Eq. (\ref{TmunuBound}) we have used the relation $K_{\nu
_{\sigma }}(\gamma a)=1/\bar{I}_{\nu _{\sigma }}^{(1,\sigma )}(\gamma a)$,
which directly follows from the Wronskian relation for the modified Bessel
functions in combination with Eq. (\ref{appBSeq}). In the case when several
bound states are present the sum of their separate contributions should be
taken.

The VEV of the energy-momentum tensor is the sum of the part coming from the
modes with real $k$ given by Eq. (\ref{qrbout}) and of the part coming from
the bound states given by Eq. (\ref{TmunuBound}). In order to transform the
first part we again rotate the integration contour in Eq. (\ref{qrbout}) by
the angle $\pi /2$ for $s=1$ and by the angle $-\pi /2$ for $s=2$. But now
we should take into account that the integrand has poles at $x=\pm i\gamma a$
which are zeroes of the functions $\tilde{H}_{\nu _{\sigma }}^{(s)}(e^{\eta
_{s}\pi i/2}x)$ in accordance with Eq. (\ref{appBSeq}). Rotating the
integration contour we will assume that the pole $e^{\eta _{s}\pi i/2}\gamma
a$, $s=1,2$, on the imaginary axis is avoided by the semicircle $C_{\rho
}^{(s)}$ in the right half plane with small radius $\rho $ and with the
center at this pole. The integration over these semicircles will give an
additional contribution%
\begin{equation}
\frac{-\delta _{\mu }^{\nu }}{16\pi \alpha ^{2}a^{3}r}\sum_{j=1/2}^{\infty
}(2j+1)\sum_{\sigma =0,1}\sum_{s=1,2}\int_{C_{\rho }^{(s)}}dx\frac{\tilde{J}%
_{\nu _{\sigma }}(x)}{\tilde{H}_{\nu _{\sigma }}^{(s)}(x)}f_{\sigma \nu
_{\sigma }}^{(\mu )}\left[ x,H_{\nu _{\sigma }}^{(s)}(xr/a)\right] .
\label{poles}
\end{equation}%
By evaluating the integrals in this formula it can be seen that this term
cancels the contribution (\ref{TmunuBound}) coming from the corresponding
bound state. Hence, we conclude that the formulae given above for the
core-induced parts in the VEVs are valid in the case of the presence of
bound states as well.

In the flower-pot model the equation (\ref{appBSeq}) for the bound states
takes the form%
\begin{equation}
C\{I_{l-1/2}(\alpha a\gamma ),K_{l/\alpha -1/2}(a\gamma )\}=0.
\label{FlpotBS}
\end{equation}%
In Section \ref{sec:flowerpot} we have shown that the function on the left
of this equation is always negative and, hence, in the flower-pot model no
bound states exist.

\end{document}